\documentclass[12pt]{article}
\usepackage{cite}
\usepackage{psfrag}
\usepackage{graphicx}
\usepackage{amsmath,amssymb}
\newcommand{\sech}{\textrm{sech\,}}

\begin{document}
\title{Efficient numerical method\\of the fiber Bragg grating synthesis}
\author{O.V. Belai,$^*$ L.L. Frumin,$^\dag$ E.V. Podivilov,$^*$
D.A. Shapiro,$^*$ \\ $*$ Institute of Automation and Electrometry,\\
Siberian Branch, Russian Academy of Sciences,\\
1 Koptjug Ave, Novosibirsk, 630090 Russia; \\
$\dag$ Novosibirsk State University,\\ 2 Pirogov St., Novosibirsk
630090, Russia}

\maketitle

\begin{abstract}
A new numerical method is developed for solution of the Gel'fand --
Levitan -- Marchenko inverse scattering integral equations. The
method is based on the fast inversion procedure of a Toeplitz
Hermitian matrix and special bordering technique. The method is
highly competitive with the known discrete layer peeling method in
speed and exceeds it noticeably in accuracy at
high reflectance.
\end{abstract}

\section{Introduction}

Promising technological applications of fiber Bragg gratings (FBG)
\cite{A02} stimulate research and development of numerical methods
of their synthesis. The propagation of counter-directional waves in
single-mode fiber with quasi-sinusoidal refractive index modulation
is described by coupled wave differential equations \cite{RK99}.
Calculation of reflection coefficient $r(\omega)$ from given
coordinate dependence of the refractive index is the direct
scattering problem. The inverse scattering problem consists in
recovery of the refractive index from given frequency dependence of
the reflection coefficient $r(\omega)$. In mathematical physics the inverse
problem for coupled wave equations reduces to coupled Gel'fand --
Levitan -- Marchenko (GLM) integral equations \cite{jZS71}. However,
the straightforward numerical solution of the GLM equations is
usually considered as too complicated for
practical FBG synthesis. At first sight it requires $N^4$
operations, where $N$ is the number of discrete points along the
grating.

Since, the solution of integral equations seems to be inefficient,
other numerical methods of FBG synthesis are elaborated. In
particular, iterative methods with $lN^3$ operations are widespread,
where $l$ is the number of iterations necessary for convergence. For
instance, they are successive kernel approximations by Frangos and
Jaggard \cite{etapF95}, high-order Born approximations by Peral et
al \cite{i3ePCM96} or advanced algorithm by Poladian \cite{oftP99}
which uses information about the reflection characteristics from
both ends of the grating. Sometimes additional approximations are
applied. For example, Song and Shin \cite{josSS85} approximate the
reflection spectrum by a rational function or Ahmad and Razzagh
\cite{amcAR98} approximate the kernel function of integral equations
by polynomials.

The alternative approach is the layer peeling method known from
quantum mechanics and geophysics and applied for FBG synthesis by
Feced et al \cite{i3eFZM99}, Poladian \cite{olP00} and Skaar et al
\cite{i3eSWE01}. The method has a clear physical
interpretation of the reflected signal as a superposition of impulse
responses from different uniform layers or point reflectors placed
along the grating. Each thin layer has small reflectivity and can be
taken into account within the first Born approximation. Because of
high efficiency (of the order of $N^2$ operations) this method
becomes widely used. The disadvantage of conventional layer peeling
is the exponential decay of accuracy along the grating because of
error accumulation during the reconstruction process
\cite{josaaSF02}. The comparable efficiency $N^2$ was demonstrated by
Xiao and Yashiro \cite{tapXY02} who transformed the GLM integral
equations to hyperbolic set of partial differential equations and
solved it numerically. This approach have several modifications,
in particular, Papachristos and Frangos \cite{josaPF02} came to
second-order partial differential equations
and also solved them numerically.

Better results at high reflectance are demonstrated by combination
of the iterations and the layer peeling. It is the integral layer
peeling method proposed by Rosenthal and Horowitz \cite{i3eRH03}.
The grating is divided into thin layers, but layers are not assumed
to have uniform profile. The profile of each layer is found by
iterative solution of GLM equations.

A recent attempt of straightforward numerical solution was made in
\cite{josabBFPSS06}. The GLM equations was solved with the help of a
bordering procedure and Cholesky decomposition. This approach takes
of the order of $N^3$ operations. The aim of present paper is to
propose more efficient numerical algorithm with $O(N^2)$ operations. The
improvement is possible due to specific symmetry of the matrix in
the discrete GLM equations, the Toeplitz symmetry: the elements of
any one diagonal are the same. The Toeplitz symmetry leads to
considerable decrease in the number of operations, similar to the
fast algorithms by Levinson \cite{jmpL47}, Trench \cite{jsiamT64}
and Zohar \cite{jacmZ74}. The proposed method utilizes a modified
bordering procedure and a second-order approximation of integrands,
the Hermitian symmetry is also taken into account.

The paper is organized as follows. In Sec.~\ref{sec:2} the GLM
equations are reduced to convenient form for numerical calculation.
The algorithm based on the specific ``inner-bordering'' technique
and Toeplitz symmetry is described in Sec.~\ref{sec:3}. Testing
numerical calculations and their comparison with the generalized
hyperbolic secant (GHS) exactly solvable profile and discrete layer
peeling (DLP) results are summarized in Sec.~\ref{sec:4}.

\section{GLM equations}\label{sec:2}

Let us consider the propagation of light through a grating with
refractive index $n+\delta n(x)$ consisting of homogeneous
background $n=\textrm{const}$ and weak modulation $\delta n \ll n$.
The refractive index modulation is quasi-sinusoidal
$$
\delta n(x)/n=2\alpha(x)\cos\left(\kappa
x+\theta(x)\right),
$$
where $\kappa$ is the spatial frequency, $\alpha(x)$ is the apodization
function \cite{RK99} and $\theta(x)$ is the phase modulation that
describes the chirp of the grating, variation of its spatial
frequency. These functions are supposed to be slow-varying, that is,
$\alpha' \ll \kappa\alpha$, $\theta' \ll \kappa$, where prime
denotes the coordinate derivative. The detuning
$\omega=k-{\kappa}/{2}$ of the light wave with respect to the
grating resonance frequency $k_0=\kappa/2$ is supposed to be small,
$\omega\ll\kappa/2$. The wave propagation can be described by the
coupled wave equations:
\begin{equation}\label{CWE}
 \psi_1' - i \omega \psi_1= q^\ast \psi_2, \quad
 \psi_2' + i \omega \psi_2= q\; \psi_1,
\end{equation}
where asterisk denotes the complex conjugation,
the coupling coefficient  $q(x)$ is defined by
$q(x)=-i\alpha(x) k_0 e^{-i\theta(x)}$.

The inverse problem for coupled wave equations was studied by
Zakharov and Shabat \cite{jZS71}, see also \cite{josSS85}.
The problem was reduced to the Gelfand --- Levitan --- Marchenko
coupled integral equations
\begin{eqnarray}\nonumber
    A_1(x,t) + \int\limits^{x}_{-\infty}  R(t+y) A^{*}_{2}(x,y)\,dy =0, \\
    A_2(x,t) +\int\limits^{x}_{-\infty}  R(t+y) A^{*}_{1}(x,y)\,dy = - R(x+t),
    \label{GLM}\\x>t.\nonumber
\end{eqnarray}
Here
\begin{equation}
R(t)= \frac{1}{2 \pi} \int\limits_{-\infty}^\infty
r(\omega)e^{-i\omega t}\,d\omega\label{r_of_t}
\end{equation}
is the Fourier transform of the left reflection coefficient
$r(\omega)$. For finite grating in the interval $0\leqslant x
\leqslant L$ kernel functions $A_{1,2}(x,t)$ are not equal to zero
only within triangular domain $-x<t<x$. Due to the causality principle the
impulse respond function equals zero $R(t) = 0$ at $t<0$. Integral
equations (\ref{GLM}) are closed in triangular domain $-x<t<x<L$ and
allow one to find the kernel functions $A_{1,2}(x,t)$
from function $R(t)$ given in interval $0<t<2L$. The complex
coupling coefficient $q(x)$ can be found from the synthesis relation
\begin{equation}\label{synth}
    q(x)=2\lim\limits_{t\to x-0} A_2(x,t).
\end{equation}

For numerical analysis let us introduce more convenient variables
$u(x,s)= A^{*}_{1}(x,x-s)$, $v(x,\tau)= A_{2}(x,\tau-x)$. GLM
equations (\ref{GLM}) take the form
\begin{eqnarray}
    u(x,s) + \int\limits^{2x}_{s}  R^{*}(\tau-s) v(x,\tau)\,d\tau =0,\nonumber\\
    v(x,\tau) +\int\limits^{\tau}_{0} R(\tau-s) u(x,s)\, ds = - R(\tau). \label{uvGLM}
\end{eqnarray}
Functions $u(x,\tau)$, $v(x,\tau)$ are determined in domain
$0\leqslant\tau\leqslant 2x\leqslant2L$. The synthesis relation
(\ref{synth}) can be rewritten as
\begin{equation}\label{uv-synth}
    q(x)=2v(x,2x-0).
\end{equation}
The integral operator in equations (\ref{uvGLM}) acting in the space
of two-component vectors constructed from functions $u$, $v$ is
Hermitian. Note that function $R$ in integrands of Eq. (\ref{uvGLM}) depends on difference of
variables only. This property resulting in Toeplitz symmetry of the matrix obtained
by discretization of integral operator is exploited in the next section.

\section{Numerical procedure}\label{sec:3}

For numerical solution of Eq. (\ref{uvGLM}) let us consider their
discrete analogue. Divide interval $0\leqslant\tau\leqslant2L$,
where function $R(\tau)$ is known, by segments of length $h = 2L/N$.
Introduce the discrete variables $\tau_n$, $s_k$, $x_m$ in
accordance with
\begin{eqnarray}\label{eq5}
    s_k=h\left(k-\frac{1}{2}\right),\quad k=1,\dots, m,\nonumber \\
    \tau_n=h\left(n-\frac{1}{2}\right), \quad n=1,\dots, m,\\
    x_m = \frac{m h}{2}, \quad m = 1,\dots, N.\nonumber
\end{eqnarray}

Define grid functions $u^{(m)}_{n} =u(x_m,\tau_n)$,
$v^{(m)}_{n}=v(x_m,\tau_n)$ and $R_{n} = R(hn)$. The integrals in
(\ref{uvGLM}) can be approximated by the simplest rectangular
quadrature scheme or more accurate trapezoidal scheme thus being
transformed into sums. The accuracy of the algorithm for rectangular
approximation is $O(N^{-1})$, for trapezoidal one it is $O(N^{-2})$.

Discrete form of GLM equations for rectangular approximation is
\begin{eqnarray}
    u^{(m)}_{k} +h\sum_{n=k}^{m}  R^*_{n-k} v^{(m)}_{n} = 0,\nonumber\\
    v^{(m)}_{n} +h \sum_{k=1}^{n}  R_{n-k} u^{(m)}_{k} = -R_{n},\label{eq6}\\
    n,k=1,\dots, m, \;\; m=1,\dots, N.\nonumber
\end{eqnarray}
The synthesis relation for the complex mode coupling coefficient
(\ref{uv-synth}) with accuracy $O(N^{-1})$ is
\begin{equation}\label{eq7}
    q^{(m)}=2v^{(m)}_{m}.
\end{equation}
The set (\ref{eq6}) at fixed index $m$ can be represented as one
matrix equation
\begin{equation}\label{eq8}
    {\bf G^{(m)} w^{(m)}}={\bf b^{(m)}},
\end{equation}
where vector ${\bf w^{(m)}}$ of dimension ${2m}$ is arranged from
the grid functions $u^{(m)}_n$ and $v^{(m)}_n$, namely,
$$
{\bf w}^{(m)}=\begin{pmatrix} {\bf u}^{(m)} \\ {\bf v}^{(m)}
\end{pmatrix}.
$$
Vector ${\bf b^{(m)}}$ is arranged from the zero
vector of dimension $m$ and the vector of dimension $m$ with
components $-R_{n}$. Square $2m\times 2m$ matrix ${\bf G^{(m)}}$  is
a block matrix
\begin{equation}\label{eq9}
    {\bf G^{(m)}}=
    \begin{pmatrix}
    {\bf E} & h{\bf R}^{\dag} \\
    h{\bf R} & {\bf E} \\
    \end{pmatrix}.
\end{equation}
Here ${\bf E}$ is the identity $m\times m$ matrix, ${\bf R}={\bf R}^{(m)}$
is the lower triangular Toeplitz $m\times m$ matrix of the form
\begin{equation}\label{eq10}
    {\bf R}=
\begin{pmatrix}
    R_0 & 0 & 0 & \dots & 0\\
    R_1 & R_0 & 0 & \dots & 0\\
    R_2 & R_1 & R_0 & \dots & 0\\
    \vdots &\vdots &\vdots &\ddots & 0 \\
    R_{m-1} & R_{m-2} & R_{m-3} & \dots & R_{0}
\end{pmatrix}.
\end{equation}
Matrix ${\bf R}^{\dag}$ is the upper triangular Toeplitz $m\times m$
matrix, that is Hermitian conjugate to matrix ${\bf R}$.
Block matrix ${\bf G^{(m)}}$ is also Toeplitz and Hermitian.

The solution of the algebraic set (\ref{eq8}) can be found by the
inversion of matrix ${\bf G^{(m)}}$ using, for example, the Levinson
bordering algorithm \cite{jmpL47}. However, we should fulfill much
simpler task of finding complex mode coupling coefficient $q^{(m)}$
with the help of (\ref{eq7}) which requires only the lower element
of vector $w^{(m)}_{2m}=v^{(m)}_{m}$ to be known. Then the lower row
of inverse matrix $\left({\bf G^{(m)}}\right)^{-1}$ is interesting
for us first of all. It is known that the inverse matrix to Toeplitz
matrix is generally not Toeplitz, but it is persymmetric, i.e.,
symmetric with respect to the secondary diagonal \cite{B85}.
Therefore, its lower row is the reflection of its left column
$$
  \mathbf{f}^{(m)}=
  \begin{pmatrix}
    {f}^{(m)}_1 \\ \vdots \\ f^{(m)}_{2m}
  \end{pmatrix}
$$
with respect to its secondary diagonal. The left column in its turn
satisfies the relation
\begin{equation}\label{eq11}
    {\mathbf G}^{(m)}{\mathbf f}^{(m)}=
    \begin{pmatrix}
    1 \cr 0 \\ \vdots \\ 0
    \end{pmatrix}.
\end{equation}
The vector-column in the right hand side of (\ref{eq11}) is the first
column of the identity matrix $2m\times 2m$.

Let us also account for the Hermitian symmetry of matrix ${\mathbf G}^{(m)}$.
As known, the matrix inverse to Hermitian is also Hermitian. Owing to
persymmetry and hermicity of inverse matrix its right column is
$$
\tilde{{\mathbf f}}^{(m)} =
\begin{pmatrix}f^{*(m)}_{2m} \\ \vdots \\ f^{*(m)}_{1}
\end{pmatrix}.
$$
Tilde denotes hereafter the inverted numeration of components
along with the complex conjugation. The right column of the
inverse matrix satisfies the relation
\begin{equation}\label{eq12}
    {\mathbf G^{(m)}} \tilde{{\mathbf f}}^{(m)}=
    \begin{pmatrix}
    0 \\ \vdots \\ 0  \\ 1
    \end{pmatrix}.
\end{equation}
The last column of the identity matrix enters the
right hand side.

Since the unknown vector ${\bf w}^{(m)}$ is formed from two
vectors of dimension $m$, it is convenient for us to present the left
column of the inverse matrix ${\mathbf f}^{(m)}$ as a merging of
two vectors of dimension $m$:
$$
{\mathbf f}^{(m)} =
\begin{pmatrix}
   {\bf y}^{(m)} \\ {\bf z}^{(m)}
\end{pmatrix}.
$$

The same relations, (\ref{eq11}) and (\ref{eq12}), are valid
for left column ${\mathbf f}^{(m+1)}$ and right column
$\tilde{{\mathbf f}}^{(m+1)}$ of the inverse matrix
$\left({\mathbf G^{(m+1)}}\right)^{-1}$ at the next $(m+1)$-th step.

Similar to Levinson's algorithm \cite{jmpL47}, vectors
${\bf y}^{(m+1)}$ and  ${\bf z}^{(m+1)}$ at the next $(m+1)$-th step
can be found by means of a bordering procedure from the vectors known at
the previous $m$-th step
\begin{eqnarray}\label{eq13}\nonumber
    {\bf y}^{(m+1)} = c_m
    \begin{pmatrix}
{\bf y}^{(m)} \\ 0
     \end{pmatrix} + d_m
     \begin{pmatrix}
     0 \\ \tilde{{\bf z}}^{(m)}
     \end{pmatrix}, \\
{\bf z}^{(m+1)} = c_m
     \begin{pmatrix}
{\bf z}^{(m)} \\ 0
     \end{pmatrix} + d_m
     \begin{pmatrix}
     0 \\ \tilde{{\bf y}}^{(m)}
     \end{pmatrix}.
\end{eqnarray}
Note that the compound structure of the vectors is just what makes
the bordering procedure ``inner'', since extending vectors ${\bf y}^{(m)}$,
${\bf z}^{(m)}$ by zeros means inserting of two rows and two columns
into matrix ${\mathbf G^{(m)}}$ with one row and one column placed in
the middle of the matrix.
At the first step we find from $2\times2$ matrix ${\bf G}^{(1)}$
that
$$
y^{(1)}_1=\frac{1}{1-h^2|R_0|^2},\quad
z^{(1)}_1=\frac{-hR_0}{1-h^2|R_0|^2}.
$$

Unknown coefficients $c_m$, $d_m$ can be obtained from
relations (\ref{eq11}) and (\ref{eq12})
\begin{eqnarray}\label{eq14}
    c_m = \frac{1}{1 - | \beta^{(m)}|^2} \quad
    d_m = \frac{\beta ^{(m)}}{1 - |\beta ^{(m)}|^2},
\end{eqnarray}
with coefficient $\beta^{(m)}$ computed by formula
\begin{equation}\label{eq15}
    \beta^{(m)}=-h\left( R_{m} y^{(m)}_1+R_{m-1}
    y^{(m)}_2+\dots+R_{1}y^{(m)}_{m} \right).
\end{equation}
Then the last component $v^{(m+1)}_{m+1}$ of vector ${\bf w}^{(m+1)}$
is calculated as the convolution of the last row of the inverse matrix with
right hand side ${\bf b}^{(m+1)}$.

Actually, the last convolution is excessive, since relation
$q^{(m+1)}=2\beta^{(m+1)}/h$ holds. Thus, the number of arithmetic
operations at each $(m+1)$-th step is of the order of $m$. Then the
total number of required operations is $N^2$ which is approximately
the same as in DLP method.

A great advantage of the new algorithm appears when we use the
trapezoidal rule \cite{NR-92}, i.e., the piecewise linear
approximation of functions. The equations in this case remain
unchanged except of the right-hand side in (\ref{eq6}) that should
be replaced by $-(R_n+R_{n-1})/2$ and the main diagonal of matrix
${\bf R}$ in (\ref{eq10}) that should be given with weight 1/2.

Since the new procedure is based on Toeplitz symmetry of the matrix and
the specific procedure putting a column and a row
inside the matrix, we call it Toeplitz inner bordering (TIB) method.

\section{Testing examples}\label{sec:4}

\begin{figure}\centering
\psfrag{Freq}[(0,-10)][0][1][0]{Frequency,
$10^{-4}\omega_0$}\psfrag{Refl}[(0,0)][0][1][0]{Reflectance, dB}
\includegraphics[width=0.75\textwidth]{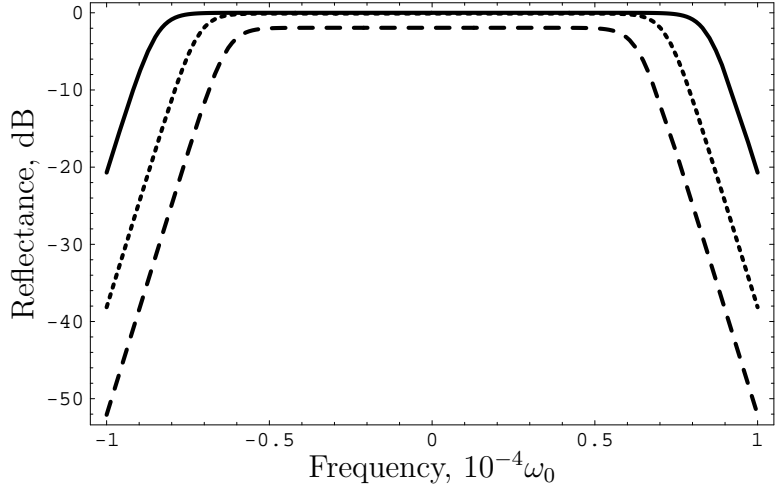}
\caption{The reflection spectrum of GHS grating for testing examples,
$k_0\mathcal{L}=5\times 10^4$, $\mathcal{F}=3$, $\mathcal{Q}=$ 1
(dashed line), 2 (dots), 3 (solid).}
\label{subfig:1}
\end{figure}

\begin{figure}\centering
\psfrag{Freq}[(0,10)][0][1][0]{Frequency,
$10^{-4}\omega_0$}\psfrag{GD}[(0,10)][0][1][0]{Group delay, $10^3$
ps}\includegraphics[width=0.75\textwidth]{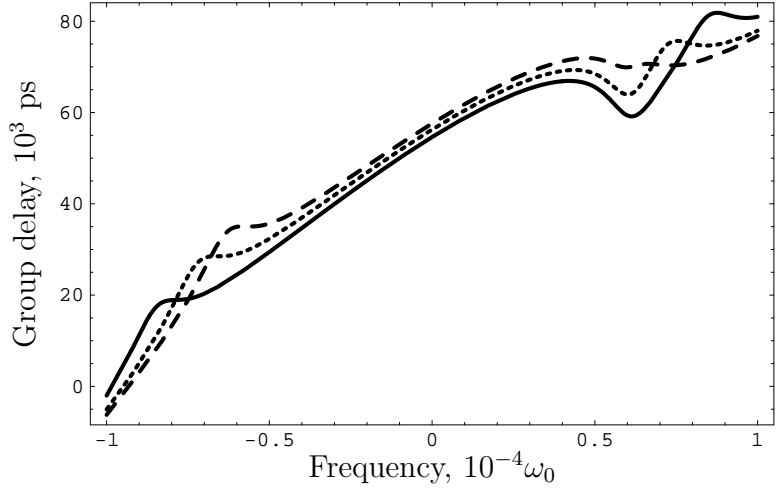}
\caption{The group delay characteristics of GHS spectrum
for testing examples at the same parameters,
as in Fig.~\ref{subfig:1}.}
 \label{fig:hyperbolic}
\end{figure}

The new method is tested using a specific case of the family of
exactly solvable chirped GHS profile of the coupling coefficient \cite{jopaPST06}
\begin{equation}\label{secant-gen}
q(x)= \frac {\mathcal{Q}}{\mathcal{L}}
\left(\sech\frac{x}{\mathcal{L}}\right)^{1-2i\mathcal{F}}.
\end{equation}
It describes a FBG with apodization function
\begin{equation}\label{apodization}
\alpha(x)=\frac{\delta n_{\max}}{2 n }\sech\frac{x}{\mathcal{L}}
\end{equation}
and phase modulation
\begin{equation}\label{chirp}
 \theta(x) = 2\mathcal{F} \ln
\left(\cosh\frac{x}{\mathcal{L}}\right) -\frac\pi2,
\end{equation} where
$\mathcal{L}$ is the half width of grating apodization profile at
level $\sech(1)=0.648$, parameter $\mathcal{Q} = {\kappa \mathcal{L}
\delta n_{\max}}/{4n}$ is the grating strength (the number of
strokes through length $\mathcal{L}$ multiplied by the modulation
depth of the refractive index). Parameter $\mathcal{F}$ describes
the value of the chirp: the profile has a slowly varying spatial
frequency
\begin{equation}\label{spatial}
\kappa (x) = \kappa +\frac{d\theta}{dx} =\kappa +
\frac{2\mathcal{F}}{\mathcal{L}} \tanh\frac{x}{\mathcal{L}},
\end{equation}
that goes smoothly from one constant spatial frequency
$\kappa - {2\mathcal{F}}/{\mathcal{L}}$ to another
$\kappa + {2\mathcal{F}}/{\mathcal{L}}$.

The coupled wave equations (\ref{CWE}) have
an exact solution that can be expressed via the Gaussian hypergeometric
function. It gives the reflection coefficient of the form \cite{jopaPST06}
\begin{equation}
\label{eq:rqfinal2} r(\omega) = -2^{-2i\mathcal{F}}\mathcal{Q}
\frac{\Gamma(d)}{\Gamma(d^*)} \frac{\Gamma(f_-)} {\Gamma(g_-)}
\frac{\Gamma(f_+)} {\Gamma(g_+)},
\end{equation}
where arguments of Euler gamma-function \cite{BE53} are given by
relations:
\begin{eqnarray*}
d=\frac12+i\left[\omega\mathcal{L}-\mathcal{F}\right],\nonumber\\
f_\pm=\frac12-i\left[\omega\mathcal{L} \pm
\sqrt{\mathcal{F}^2+\mathcal{Q}^2}\right],\nonumber\\
g_\pm=1-i\left[\mathcal{F}\pm
\sqrt{\mathcal{F}^2+\mathcal{Q}^2}\right].
\end{eqnarray*}
The reflection spectrum is expressed in terms of elementary
functions
\begin{equation}\label{reflectance}
|r(\omega)|^2=\frac{\cosh
2\pi\sqrt{\mathcal{Q}^2+\mathcal{F}^2}-\cosh 2\pi\mathcal{F}}{\cosh
2\pi\sqrt{\mathcal{Q}^2+\mathcal{F}^2}+\cosh 2\pi\omega\mathcal{L}}.
\end{equation}

For numerical calculations we choose gratings with
$\mathcal{L}=5\times10^{4}/k_0, \mathcal{F}=3$ and
$\mathcal{Q}=1,2,3$, where $k_0=2\pi n/\lambda_0$ and the central
resonance wavelength is $\lambda_0=1.5\;\mu$m. Their maximum
reflectances, $|r|^2=0.6393, 0.9777, 0.9996$, are referred hereafter as
small, medium and high respectively. The reflection spectrum calculated by
formula (\ref{reflectance}) is shown in Fig.~\ref{subfig:1}. The
frequency detuning from resonance is shown in units
$10^{-4}\omega_0$, where $\omega_0$ is the central frequency of the
reflection spectrum. The spectrum is quasi-rectangular with flat top
inside the Bragg reflection band. The reflectance increases with
optical strength parameter $\mathcal{Q}$. The width of the band
$\Delta\omega\simeq 2\sqrt{\mathcal{Q}^2+\mathcal{F}^2}/\mathcal{L}$
increases, too. The group delay characteristics are plotted in
Fig.~\ref{fig:hyperbolic}. Each curve is close to straight line within the band
except of the band edges.

\begin{figure}\centering
\psfrag{abs}{$\log_2 \frac{1}{N}$}
\psfrag{ord}{$\log_2\sigma$}\psfrag{x}{$\times10$}
\includegraphics[width=0.75\textwidth]{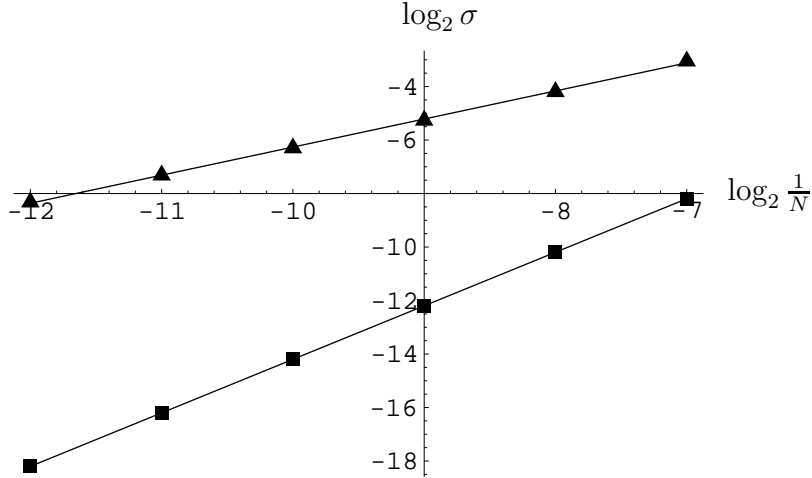}
 \caption{Root mean square error $\sigma$ of the first-order (triangles)  and
 second-order (boxes)
 reconstruction as a function of $1/N$ in logarithmic coordinates
 for $N=128\div4096$, $\mathcal{Q}=1$,
 $\mathcal{L}=5\times10^4/k_0$, $\mathcal{F}=1$. The straight lines show the
 least-square linear fitting.}
 \label{fig:accuracy}
\end{figure}

The GLM equations for reflection coefficient (\ref{eq:rqfinal2}) are
solved using the method described in Sec.~\ref{sec:3}. As the first
step of calculations the fast Fourier transform Eq. (\ref{r_of_t})
is performed at sufficiently long frequency interval and small
frequency step $\delta\omega=2\pi/L_{max}$, where
$L_{max}=35\mathcal{L}$, in order to neglect the values outside both
the frequency and the coordinate intervals where the reflection
spectrum and the grating are defined. The frequency domain for
integration is defined as $-\Omega/2\leq\omega\leq\Omega/2)$, where
$\Omega=N_{\omega}\delta\omega$ and $N_{\omega}$ is the number of
discrete points in frequency.  While we are going to test the method
of solving GLM equations itself, the additional errors produced by
the Fourier transform should be minimized. For this purpose the
excessively precise determination of function $R(t)$ is made. In
order to provide the sufficient accuracy for the second-order method
we choose $N_{\omega}\gg N$, in particular, $N_{\omega}=2^{20}$ at
$N=2^{12}$. It does not significantly increase the total number of
operations, since the Fourier transform requires $N_{\omega}\log_2
N_{\omega}$ operations and done only once.

The inaccuracy of rectangular and that of trapezoidal quadrature
formulas are compared. Root mean square error $\sigma$ of the
grating reconstruction is shown in Fig.~\ref{fig:accuracy} as a
function of $1/N$. As evident from the figure the first and
second-order algorithms result in different errors. For the first-order
method the dependence is linear, whereas for the second order it
becomes nearly quadratic. The slopes of fitted straight lines are
1.05 and 2.00, respectively. Moreover, the error of the second order
method is significantly less at $N\geqslant 2^{6}$. Then the
second-order method is applied in all calculations below.

\begin{figure}\centering
\psfrag{ord}[-100,0]{Apodization function $\alpha\times 10^{5}$}
\psfrag{abs}[0,-20]{Coordinate $k_0x\times 10^{-5}$}
\includegraphics[width=0.75\textwidth]{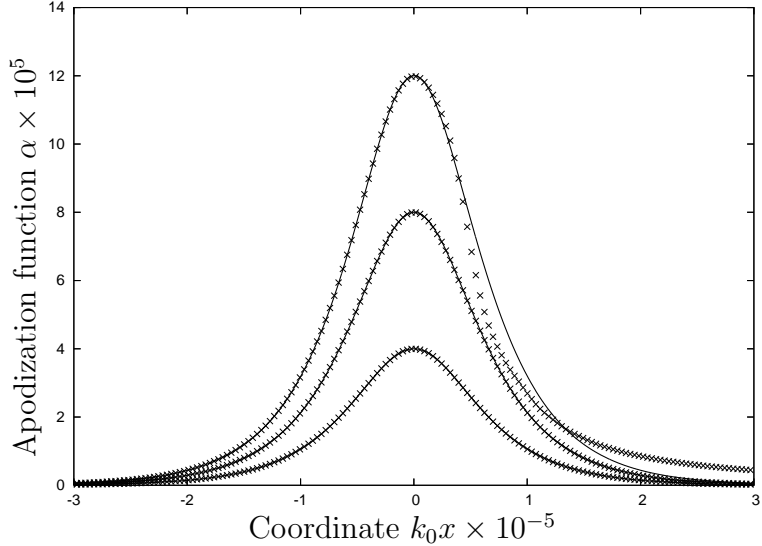} \caption{The envelope
$\alpha$ as a function of coordinate reconstructed by second-order
TIB (solid line) and by DLP (crosses): from the top down
$\mathcal{Q}=3,2,1$.} \label{fig:Compare}
\end{figure}

\begin{figure}\centering
\psfrag{abs}[0,-20]{Coordinate $k_0x\times 10^{-5}$}
\psfrag{ord}[-100,0]{Deviation $\delta\alpha\times10^8$}
\includegraphics[width=0.75\textwidth]{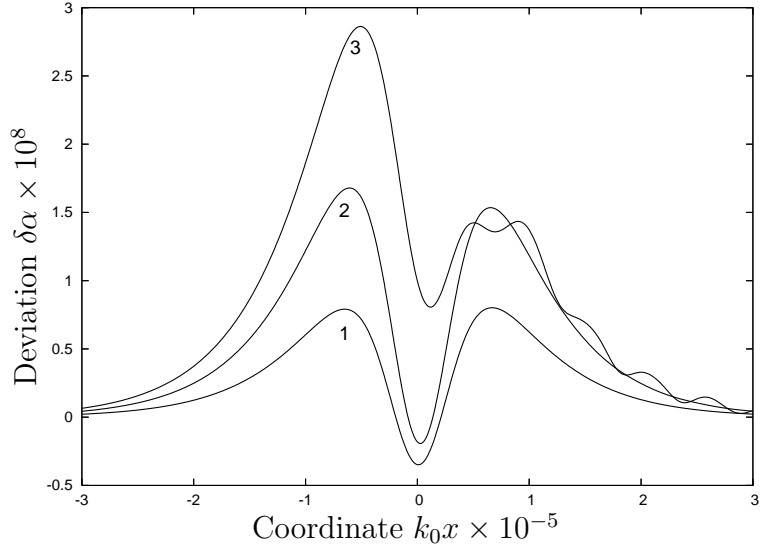}
\caption{Comparison of the second-order TIB method with GHS profile
(\protect\ref{apodization}): the deviation of numerical calculations
from the analytical formula as a function of coordinate. The number
near each curve denotes the value of grating strength
$\mathcal{Q}$.} \label{fig:Deviation}
\end{figure}

\begin{figure}\centering
\psfrag{ord}[-100,0]{$\frac{1}{k_0}\frac{d\theta}{dx}\times
10^{-5}$} \psfrag{abs}[0,-20]{Coordinate $k_0x\times 10^{-5}$}
\includegraphics[width=0.75\textwidth]{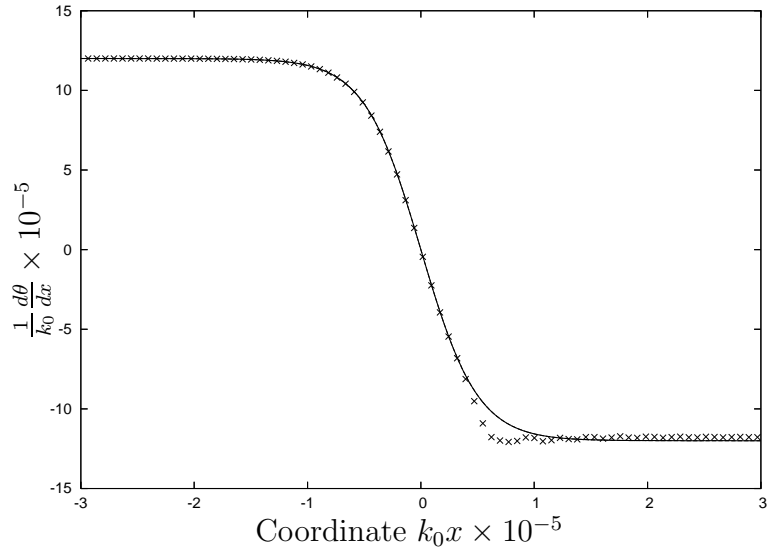}
\caption{The deviation of the spatial frequency of the grating
(\protect\ref{spatial}) from $\kappa$ calculated by TIB method
(solid line) and DLP (crosses) at
$\mathcal{Q}=3$.}\label{fig:spatial}
\end{figure}

The comparison of the second-order TIB with DLP reconstruction at
fixed $N$ reveals that TIB method occurs $2\div3$ times faster. The
apodization function $\alpha(x)$ at the same parameters, as in
Fig.~\ref{subfig:1}, and $N=8192$ is shown in
Fig.~\ref{fig:Compare}. For relatively weak grating
$\mathcal{Q}=1,2$ both methods are appropriate, as bottom curves
demonstrate, and the resultant curves are in agreement with formula
(\ref{apodization}). However, for strong grating the DLP calculation
gives significant error. The reason is probably the error
amplification in DLP \cite{josaaSF02}.

The deviation of TIB solution from GHS profile
(\ref{apodization}) is shown in Fig.~\ref{fig:Deviation}. The
deviation is maximal near the center of the profile and negligible
at the ends. The curves are regular at small and medium strength and
acquire irregular behavior for strong grating. The maximum relative
error of reconstructed apodization function is less than
$2.5\cdot10^{-4}$ for all studied parameters.

The phase characteristics of complex coupling coefficient
demonstrate the similar features. At $\mathcal{Q}=1,2$ the phase
characteristics calculated by TIB and DLP methods are close. At high
optical strength $\mathcal{Q}=3$ the error of DLP grows up towards
the right end. The spatial frequency $\theta'$ of reconstructed
profile is shown in Fig.~\ref{fig:spatial}. The smooth transition
between two horizontal asymptotes of analytical expression
(\ref{spatial}) is reproduced by TIB calculation for
$\mathcal{Q}=3$, whereas the DLP gives the deviation at the right
side of the curve.

\section{Discussion}

The discrete layer peeling \cite{i3eSWE01} calculates $q$ at the
input end of the grating and then truncates the grating dealing
every next step with shorter grating residue. This is the reason of
error accumulation throughout the calculation from the input layer
to the output one. The TIB method of matrix inversion recovers the
complex coupling coefficient $q(x)$ along the whole length at one
step. Then it has higher accuracy at comparable efficiency.

It is possible to make TIB even more efficient dividing the length
$L$ by segments. After reconstruction of the coupling coefficient in
the current segment one could find the amplitudes of opposite waves
at the input end of the next segment and repeat the procedure with
the next segment. The efficiency could be improved if we choose the
optimal number of segments.

The similar combined procedure with indirect iterative solution of
GLM equations, known as integral layer peeling (ILP), leads to fast
reconstruction of a grating \cite{i3eRH03}. In that approach the
grating is divided by $M$ layers with $m$ intermediate points in
each. The total number of points along the grating is $N=mM$. The
reconstruction problem in each layer is solved by an iterative
procedure applied to GLM integral equations. The reflection
coefficient of truncated grating after a peeling step is found with
high accuracy. The computational complexity of ILP is of the order
of
\[
n_{\rm total}\sim \left(lNm+\frac{l+1}m N^2\right)\log_2 N
\]
required operations \cite{i3eRH03}, where $l$ is the number of
iterations during the reconstruction of a layer. At $l=0$ and large
$m$ the complexity becomes less than $N^2$. However with increasing
$m$ and decreasing $l$ the accuracy goes down fast.

If we were change the ILP iterations by the proposed TIB technique
the complexity of the reconstruction within a layer would be $N\ln
N+m^2$. We obtain the total number of required operations
multiplying it by $M=N/m$:
\[
n_{\rm total}\sim\frac{N^2}m\log_2 N+mN.
\]
This number has a minimum value $\min n_{\rm total}\sim N^{3/2}(\ln
N)^{1/2}\ll N^2, N\to\infty$ at $m\sim(N\ln N)^{1/2}$. As long as
each layer is sufficiently thin and its optical strength is not large
($\mathcal{Q}\lesssim 1$), the matrix inversion method shall give
the superior result compared to iterations.

For very strong gratings at $1-|r|\to0$ all the methods lose their
accuracy, since an eigenvalue of GLM equations tends to zero and the
problem becomes ill-conditioned. If the grating is strong, then
incident light is reflected in the domain close to the input end.
Only exponentially small part penetrates far from the input end,
then it is almost impossible to reconstruct the profile of the
deeper region. Fortunately, it is a formal mathematical problem. For
more or less reasonable optical density, for instance, with maximum
reflectance up to 99.9\%, the proposed TIB method is adequately
accurate.

\section{Conclusions}

Thus the new method of the FBG synthesis is proposed. The method is
based on direct numerical solution of the coupled GLM equations. The
Toeplitz symmetry of the matrix and the inner-bordering procedure
provide fast computation, similar to known fast Levinson's
algorithm. The second-order quadrature formula sufficiently improves
the accuracy without loss of efficiency. The method is tested using
exactly solvable profile of chirped grating. The method does not
concede the DLP in speed and at the same time remains more accurate
for strong gratings.

\section*{Acknowledgment}

Authors are grateful to D. Trubitsyn and O. Schwarz for fruitful
discussions. The work is partially supported by the CRDF grant
RUP1-1505-NO-05, the Government support program of the leading
research schools (NSh-7214.2006.2) and interdisciplinary grant \# 31
from the Siberian Branch of the Russian Academy of Sciences.

\end{document}